# Multiple Access in Cellular V2X: Performance Analysis in Highly Congested Vehicular Networks

Behrad Toghi*, Md Saifuddin*, Hossein Nourkhiz Mahjoub*, M. O. Mughal*, Yaser P. Fallah*,
Jayanthi Rao†, Sushanta Das†
*Networked Systems Lab, University of Central Florida, Orlando, FL, USA
†Ford Motor Company, Dearborn, MI, USA
{toghi, md.saif, hnmahjoub, ozair}@knights.ucf.edu, yaser.fallah@ucf.edu, {jrao1, sdas30}@ford.com

*Abstract*— Vehicle-to-everything (V2X) communication enables vehicles, roadside vulnerable users, and infrastructure facilities to communicate in an ad-hoc fashion. Cellular V2X (C-V2X), which was introduced in the 3rd generation partnership project (3GPP) release 14 standard, has recently received significant attention due to its perceived ability to address the scalability and reliability requirements of vehicular safety applications. In this paper, we provide a comprehensive study of the resource allocation of the C-V2X multiple access mechanism for high-density vehicular networks, as it can strongly impact the key performance indicators such as latency and packet delivery rate. Phenomena that can affect the communication performance are investigated and a detailed analysis of the cases that can cause possible performance degradation or system limitations, is provided. The results indicate that a unified system configuration may be necessary for all vehicles, as it is mandated for IEEE 802.11p, in order to obtain the optimum performance. In the end, we show the inter-dependence of different parameters on the resource allocation procedure with the aid of our high fidelity simulator.

## I. INTRODUCTION

Over the last two decades, the quest for efficient intelligent transportation systems (ITSs), and eventually safer, greener, and smarter roads led to the introduction of the dedicated short range communication (DSRC) based on IEEE 802.11p, IEEE 1609.X, and SAE J2945 standards [1], [2]. Despite the fact that DSRC can be considered as the current primary solution for the vehicular communication, many contributors, e.g., automakers, tier-1 suppliers, and regulators, have shown interest in the Long Term Evolution (LTE) technology as an alternative solution for the vehicular ad-hoc networks (VANETs) to enable vehicles to share their position and mobility information in the form of either base safety messages (BSMs) or model-based communication, as recently proposed in [4] and [5].

The 3rd Generation Partnership Project (3GPP) announced the release 14 standard in 2016 which introduced four communication modes. Mode-1 and mode-2, which are often referred as the device-to-device (D2D) communications or proximity services (ProSe), are inherited from the earlier release 12 and define a new communication interface known as the sidelink or PC5 interface. Sidelink communication enables user equipments (UEs) to bypass the central LTE base station, i.e., eNodeB, and communicate in a peer-to-peer manner which can be utilized by multiple applications, e.g., in-door content sharing, network relaying, and low-power consumption networks. The D2D communication is not able to satisfy the strict vehicular safety requirements, especially in terms of communication reliability and latency.

In order to cater the aforementioned specifications for the vehicle-to-everything (V2X), mode-3 and mode-4 communication were specified in 3GPP release 14 standards [3]. A wide range of configurable parameters and features were defined in the release 14 which enabled designers to improve and optimize the network performance according to the specific scenarios. Enhancements in the collision avoidance, channel access, sub-channelization schemes, and hybrid automatic repeat request (HARQ) were made in order to mitigate the performance degradation and provide improved scalability, which is essential in congested and high-density vehicular scenarios. Moreover, two extra demodulation reference symbols (DMRS) were added to each subcarrier in order to provide support for the high-speed mobility use cases and compensate for the Doppler spread.

UEs operate in mode-3 when the network coverage is available and eNodeB will be responsible for scheduling and allocating resources; on the contrary, mode-4 is defined to support the out-of-coverage or partial coverage communication where resource allocation required to be in a distributed and unsupervised fashion. In contrast with the random resource allocation scheme in mode-2, which suffers from severe scalability and packet collision issues, mode-4 enjoys an enhanced packet collision avoidance mechanism which utilizes the channel occupancy record to cut down the collision probability. In the remainder of this text, our emphasis will be on the mode-4 sidelink communication, which we refer to as C-V2X and is also known as LTE-V or LTE V2V in the scientific literature.

C-V2X is a relatively recent cellular technology enhancement; hence, there is sparse scientific literature available in this topic. Among the few research articles, one can refer to system level evaluation in [7] which investigates different types of the transmission error such as propagation errors, packet collision, and errors due to the half-duplex operation. In [8], the authors present a tutorial of the C-V2X and future trends such as congestion control and release 16 developments in addition to a concise comparison versus IEEE 802.11p technology. In [9], after an introductory overview of the applications and use cases of the V2X communication, a link level performance comparison of C-V2X and DSRC is presented which demonstrates that C-V2X out-performs DSRC in most of the vehicular scenarios, e.g., highway and urban scenarios. Recent results in [10], study how the resource reservation periodicity and the number of available radio resources can affect the performance of the C-V2X communication. Results demonstrate that the communication reliability can be improved utilizing more radio resources.

In this article, we present a detailed assessment of C-V2X technology under high-density vehicular scenarios in order to analyze the impact of the various configurable parameters on the performance of the resource allocation procedure. We demonstrate that an optimum configuration can heavily affect the system performance. This perspective is in contrast with that of [7]-[10], where simplistic assumptions are made for the tunable parameters and the baseline performance in medium or low-density scenarios is under consideration. Moreover, the packet error rate (PER) and inter-packet gap (IPG) metrics are chosen to assess the system performance in terms of both network reliability and latency; such analysis is absent in the previously mentioned works. We implemented an event-based ns-3 simulator, which is precisely in compliance with the latest versions (June 2018) of 3GPP release 14 standards [3], [11]-[18], and employ this simulation platform to rigorously investigate the physical (PHY) and medium access control (MAC) layer mechanisms, and phenomena that can cause possible performance degradation or deadlock loops.

The rest of the paper is organized as follows. A comprehensive description of the cellular vehicle-to-everything communication and its resource allocation process is detailed in Section II. Simulation setup and environment are outlined in Section III. Analysis and results are described in Section IV before concluding the paper in Section V.

## II. Cellular Vehicle-to-everything Communication

In 1999, the U.S. Federal Communications Commission (FCC) specified 75 MHz of spectrum in 5.9 GHz band for the intelligent transportation systems (ITS) and vehicle safety applications. C-V2X is expected to operate in the ITS band and possibly co-exist with DSRC. In this section, we exploit different aspects of the C-V2X communication with a focus on the resource allocation and channel access mechanisms.

### A. Principles of the Sidelink Communication

New interfaces, which were introduced for the ProSe applications in 3GPP release 12 standard, are being utilized by C-V2X as well. This remodeled network architecture includes multiple additions among which Uu and PC5 links are being exploited in C-V2X. PC5 is a one-to-many communication interface which allows UEs to broadcast their messages among their neighboring groups. In mode-4, both data and control information are communicated through the PC5 link. On the other hand, Uu links a UE to the LTE air medium, known as the Evolved Universal Terrestrial Radio Access Network (E-UTRAN). In mode-3 communication, the PC5 link bears data packets and Uu is utilized for the control information exchange between a UE and the central base station. The LTE sidelink exploits four communication channels as listed in Table I.

A detailed description of the LTE generic frame structure in PHY layer is crucial in order to competently investigate and evaluate the C-V2X procedures. We follow the terminology and parameters defined in 3GPP in order to help the reader to trace the standardized documentation and implementations. A given LTE physical channel is divided into smaller fragments, both in time and frequency, which are referred to as *frames*. Every LTE frame is a 10 ms wide and its length is equal to the system bandwidth.

TABLE I.  LTE SIDELINK COMMUNICATION CHANNELS

| Ch. Level | Shared Channel | Discovery Channel | Broadcast Channel | Control Channel |
|---|---|---|---|---|
| *Logical Ch.* | STCH | - | SBCCH | - |
| *Transport Ch.* | SL-SCH | SL-DCH | SL-BCH | - |
| *Physical Ch.* | PSSCH | PSDCH | PSBCH | PSCCH |

A frame breaks down into 10 *subframes* in the time domain, i.e. each subframe is 1 ms wide and contains two *time-slots*. A time-slot (slot) is a time-series of 7 SC-FDMA *symbols* (assuming normal cyclic prefix length). Analogous segmentation is conducted for the frequency domain of the LTE frame; it subdivides into *subcarriers*, with 15 kHz spacing. A 2-dimensional time-frequency entity can be considered as a radio resource in the single-carrier frequency-division multiple access (SC-FDMA) context. Following the introductory divisions, radio resources can be investigated; a *resource element* (RE) essentially covers one symbol in time and 1 subcarrier in frequency domain. A *resource block* (RB) consists of 12 subcarriers in frequency and 1 slot, i.e., 7 symbols, in the time domain. Finally, two consecutive RBs, in the time domain, form a *scheduling block* (SB). To encapsulate the aforementioned structure, a frame with 10 MHz bandwidth contains 50 RB-lengths in frequency (1 MHZ guard-band) and 20 RB-widths in the time domain.

As described above, a subframe consists of 14 orthogonal symbols per subcarrier. Release 14 standard introduced an enhanced pilot signaling in order to support the high-speed requirements of C-V2X. Two supplementary pilot symbols, known as the de-modulation reference symbols (DMRSs), were added to the original 2 symbols defined in LTE D2D. Four DMRSs occupy the third, sixth, ninth, and twelfth symbols of each subcarrier [16]. In addition, the last symbol of each subcarrier is reserved for the guard band and the first symbol is dedicated to the automatic gain control (AGC). The AGC mechanism is not permanently operational and it is up to UE implementation to how and when activate it. We follow the 3GPP standard [18] and assume that 9 SC-FDMA symbols per subcarrier are available for data transmission.

UEs broadcast their BSMs as data blocks via the physical sidelink shared channel (PSSCH), as introduced in Table I, and utilize the same communication channel to receive the data blocks. In this text, we refer to data transmissions as transport blocks (TBs). A TB has to be transmitted in $N_{\text{PSSCH}}^{\text{RB}}$ contiguous resource blocks, per time-slot. The number of resource blocks required to transmit a TB, is a function of the data packet size, modulation order, and code rate. In addition, every TB is accompanied by the sidelink control information (SCI), broadcasted in the physical sidelink control channel (PSCCH). SCI contains the crucial information required for successful reception and demodulation of its corresponding TB and always occupies two contiguous resource blocks per time-slot. UE must transmit TB and its corresponding SCI in the same subframe. However, a TB and its SCI can be either adjacent or non-adjacent [3].

In the PHY layer, every subframe breaks into $N_{\text{subCH}}$ smaller partitions, known as *sub-channels*. Each sub-channel consists of $N_{\text{subCHsize}}$ consecutive physical resource blocks (PRBs). The set of all available sub-channels is known as the PSSCH resource pool. UE also defines a set of resources for SCI transmissions, referred to as the PSCCH resource pool. Each PSCCH resource consists of 2 contiguous PRBs. Two contrasting schemes are defined by 3GPP for PSSCH and PSCCH resource pool configuration: *(i)* TB and SCI must be placed in an adjacent fashion; *(ii)* Non-adjacent and separated resources can be allocated for TB and SCI. It should be noted that the total number of sub-channels, $N_{\text{subCH}}$, is a function of the channel bandwidth and sub-channel size, e.g., a 10 MHz channel contains 50 RBs (assuming 1 MHZ guard band) that can form five sub-channels of size 10. All allowed values for the number of sub-channels and the sub-channel size are strictly defined in [12].

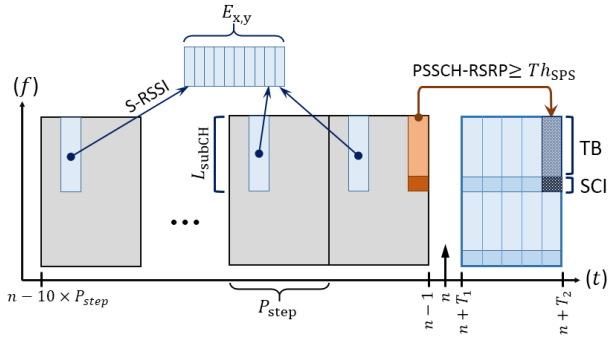

Fig. 1. Candidate single-subframe resource exemption in the sensing procedure (TB and SCI are arranged in the adjacent scheme)

The amount of data that can be transported by a given number of PRBs depends on the modulation order and code rate; C-V2X standard currently supports 16-QAM (Quadrature Amplitude Modulation) and QPSK (Quadrature Phase-shift Keying) modulations which can be used with different code rate values. The combination of a modulation order value alongside with a code rate is known as the modulation and coding scheme (MCS) index in the literature. The MCS indices defined for the LTE uplink channel, can be reused for C-V2X, however, it should be noted that modulations orders higher than 16-QAM are not supported by the current standard. The exact TB size calculation procedure is defined in [3] and will be investigated in the next sections.

### B. C-V2X Resource Allocation

C-V2X, inherently, employs the SC-FDMA, which enables a UE to access to radio resources both in time and frequency domain, i.e., two degrees of freedom. As discussed earlier, in the sidelink mode-4 communication, radio resources are allocated autonomously and in a stand-alone fashion. No acknowledgment system is designed for the sidelink communication, thus, both modes utilize blind re-transmission of redundant versions for each generated message. MAC layer of C-V2X supports the hybrid automatic repeat request (HARQ) process, restricted to two blind retransmissions [11].

In contrast with the sidelink mode-2 communication, where each UE selects resources completely randomly, mode-4 employs an enhanced resource allocation mechanism known as the sensing-based semi-persistent scheduling (SB-SPS). It is worth mentioning that the distributed resource allocation in C-V2X is eventually a random process, however, SB-SPS shrinks the available resources and significantly decreases the collision probability by limiting every UE to select resources from a narrowed-down set. The SB-SPS mechanism relies on two main concepts; first, it reduces the probability of the case that multiple UEs select a common resource and second, stochastically decouples UEs by adding randomness to the resource allocation process.

The SB-SPS mechanism can be divided into three processes: sensing, reservation, and transmission. Every UE listens to the communication channel and keeps the track of all received signals from its neighboring UEs (sensing); this record is utilized to obtain a shrunk set of resources to be reported to the MAC layer. MAC layer reserves radio resources in a semi-persistent manner by using a randomly selected counter (reservation). PHY layer assigns the semi-persistently selected physical resources to the data and control information and transmits the generated beacon to the air interface (transmission).

*1) Sensing*

Sensing procedure is illustrated in Figure 1 and discussed in this sub-section. The resource allocation granularity in C-V2X is 1ms, i.e., one subframe, which means that from the PHY layer's perspective, the smallest entity that can be allocated is one RB pair. However, higher layers take the sets of sub-channels as the smallest allocable resources. In other words, if higher layers request $L_{\text{subCH}}$ sub-channels for transmission in the subframe $y$, a candidate single-subframe resource (CSR), $R_{x,y}$, is defined as the set $\{x + j \mid j = 0, 1, \dots, L_{\text{subCH}} - 1\}$, which consists of $L_{\text{subCH}}$ consecutive sub-channels in the subframe number $y$.

Consider the case that MAC layer requests a sensing report from the lower layers at the subframe number $n$, this instance is the packet arrival time to the MAC layer. Following this request, PHY layer extracts the *sensing window* from its channel record buffer. The sensing window is defined as the set of all CSRs in $[n - 1, n - 10 \times P_{\text{step}}]$ timespan. The $P_{\text{step}}$ parameter is proposed in [23] in order to avoid synchronization conflicts, which is out of scope this paper, and is set to 100ms as suggested in [3]. Thus, PHY keeps track of all CSRs in the previous 1s, i.e., total number of $1000 \times \lfloor N_{\text{subCH}}/L_{\text{subCH}} \rfloor$ CSRs.

We introduce the *report window* as the set of all CSRs between the time frame $[n + T_1, n + T_2]$. The time offset $T_1$ can be set to any value less than or equal to 4 subframes and is preset by the higher layer depending on the required process time of the device. The maximum allowed latency, $20 \leq T_2 \leq 100$, is also preset based on the application, e.g., in the vehicular community there is a consensus on $T_2 = 100$ ms for the safety applications. PHY layer conducts an exemption procedure in order to remove CSRs with the higher likelihood of causing collisions from the report window. The narrowed down set will be reported to the higher layers to initiate the reservation process.

PHY layer initializes the set $S_A$ with all available CSRs in the report window and enforces the following exemption procedure in order to remove the likely-to-collide CSRs from $S_A$. All $R_{x,y}$ that meet at least one of the following conditions shall be excluded from the $S_A$:

*Condition 1:* Subframe $z$ in the sensing window has not been monitored and reservation horizon of $z$ overlaps with that of $y$.

*Condition 2:* A SCI message and its corresponding TB are received on the subframe $w$ in the sensing window. The reference signal received power (RSRP) of the TB is higher than a preset threshold value, i.e., PSSCH-RSRP $\geq Th_{\text{SPS}}$. The reservation horizon of $w$ overlaps with that of $y$. The concept of the reservation horizon is discussed in details in the section related to the reservation procedure.

After the exemption procedure, PHY checks if the remainder of $S_A$ contains at least 20% of the initial CSRs; if not, the exemption is repeated with a 3dB increase in $Th_{\text{SPS}}$ until the narrowed down set maintains the 20% requirement. On the next step, the metric $E_{x,y}$ is used to rank the CSRs in the narrowed-down $S_A$. $E_{x,y}$ is the linear average of the sidelink received signal strength indicator (S-RSSI) [24], that can be expressed as

$$E_{x,y} = \frac{1}{10} \sum_{i=1}^{10} \left[ \sum_{j=x}^{x+L_{\text{subCH}}-1} \text{RSSI}_{(y-i.P_{\text{step}}),j} \right] \quad (1)$$

Remaining CSRs are ranked based on the above metric and the top 20% of the CSRs with the lowest $E_{x,y}$ move to a second list, $S_B$. PHY layer reports the set $S_B$ to the MAC layer, to be used during the reservation process.

*2) Reservation*

MAC layer receives the set $S_B$ from lower layers and initiates the (re-)selection process if at least one of the following trigger conditions is met
- The random sidelink resource reselection counter (SLRRC) reaches zero.
- UE has not transmitted or retransmitted any packet during the last 1 second.
- UE has missed more reserved transmission opportunities than what is allowed by higher layers.
- UE could not meet the latency requirement for the previous transmission.
- The previously allocated resources are not sufficient for the incoming MAC protocol data unit (PDU).

Subsequent to detecting a trigger, MAC layer randomly selects a CSR from $S_B$ and periodically reserves $C_{resel}$ number of CSRs with the period, $P_{rsvp}$; reserved resources are considered as the transmission opportunities and the resource reservation interval, $P_{rsvp}$, is set by higher layers and its allowed values are {20, 50, 100, 200, …, 1000} milliseconds. Henceforth, a new random SLRRC is set with uniform probability in the range [5, 15] for $P_{rsvp} \geq 100$, [10, 30] for $P_{rsvp} = 50$, and [25, 75] for $P_{rsvp} = 20$. The time distance between the last reserved and the initially selected resources is considered as the "reservation horizon" which can be expressed as $P_{rsvp}(C_{resel} - 1)$ where $C_{resel} = 10 \times$ SLRRC [3].

After reserving the transmission opportunities, SLRRC decrements with each packet transmission and UE keeps the allocated set until a re-selection trigger occurs. If the SLRRC reaches to zero, the UE either keeps the previously selected resources or selects new resources with the probability $P_{resel}$, which can be preset to any of the values in {1.0, 0.8, 0.6, 0.4, 0.2}. If the UE happens to not select new resources, it keeps the previously selected transmission opportunities for $C_{resel}$ number of transmissions. MAC layer also has to schedule the HARQ retransmission opportunities and schedule the required number of resources. If the number of redundant versions is set to 2 and a sufficient number of resources is still left for the HARQ retransmission, MAC randomly allocates a second set of periodic transmission opportunities within the neighborhood of (-15, 15) subframe from the initially selected set of resources. The reservation process is demonstrated in Figure 2.

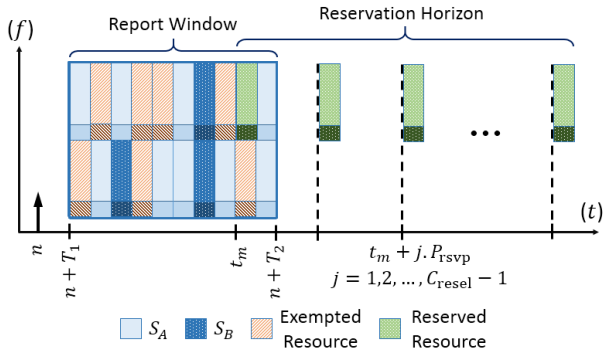

Fig. 2. Resource reservation procedure. MAC layer receives a packet at subframe number $n$ and reserves a set of periodic transmission opportunities starting from the subframe $t_m$.

TABLE II.  FOWLERVILLE PROPAGATION MODEL PARAMETERS

| Coeff. | 1-8m | 9-45m | 46-111m | 112-400m | 401-639m | >639 m |
|---|---|---|---|---|---|---|
| $\alpha$ | Linear | 1.71 | 1.77 | 1.85 | 1.88 | 1.90 |
| $m$ | 2.272 | 1.340 | 1.438 | 1.357 | 1.000 | – |

*3) Transmission*

After completing the reservation process, MAC layer instructs PHY layer about the reserved transmission opportunities. PHY layer employs the reserved CSRs to transmits the data TB in $N_{PSSCH}^{RB}$ consecutive PRBs of a reserved CSR. Moreover, PHY layer also has to load the required information to the SCI message. Neighboring UEs need this information in order to be able to successfully decode the corresponding TB. A SCI message occupies two consecutive PRBs of the reserved PSCCH resource.

### III. SYSTEM CONFIGURATION

As our main purpose is to evaluate the system level performance in vehicular scenarios involving large numbers of UEs, we implemented a link-level event-based simulator using the ns-3 simulation environment. The LENA ns-3 module, developed by Baldo et al. [21], has been employed to support the main LTE functionalities.

*A. Air Interface*

A valid and realistic physical layer realization is of great importance in a vehicular communication simulation. In order to obtain such a model, we utilized a propagation channel model from our previous work in [24] and [29] which was extracted from the field test data on the FTT-A Fowlerville Proving Ground in MI, USA by the Crash Avoidance Metrics Partnership (CAMP) consortium in collaboration with USDOT. The Fowlerville channel model was derived from a large data set of collected received signal strength indicator (RSSI) and consists of both large and small-scale propagation effects. The large-scale attenuation, $L_{LS}$, is modeled using the well-known two-ray model which can be formulated as

$$L_{LS}(d, \alpha) = 10. \alpha. \log\left(4\pi \frac{d}{\lambda} \left|1 + \Gamma e^{i\phi}\right|^{-1}\right) \text{ (dB)} \quad (2)$$

where $\alpha$ denotes the path-loss exponent, $d$ is transmitter-receiver distance, and $\lambda$ is wavelength. The reflection coefficient $\Gamma$ is a dimensionless constant and $\phi$ is the phase difference of the two interfering rays. The derived values for the $\alpha$ exponent are listed in the first row of Table II for different transmitter-receiver ranges.

To characterize the small-scale channel propagation effects, two distribution models have been chosen; a Nakagami-$m$ model, with $m$ values as denoted in Table II, for transmitter-receiver distances up to 639 m, and a Weibull distribution, with the characteristic coefficient $k = 1.4$, for the distances beyond that range.

*B. Transceiver Model*

The transmitter and receiver performance can highly impact the link budget and consequently the range of the C-V2X communication. In our simulation, we assume 9 dB noise figure [15] for the transceiver devices. Each device is equipped with 2 receive and 1 transmit Omni-directional antennas with 3 dBi gain and 1.5 m height, which is the height of a typical sedan car. The performance of the receiver side can be expressed as its ability to receive and demodulate the signals. The receiver's performance to successfully demodulate a signal depends on the received signal's signal-to-interference-and-noise ratio (SINR).

TABLE III. SIMULATION PARAMETERS

| | | | |
|---|---|---|---|
| Time ($T_{sim}$) | 100 s | Power ($P_{TX}$) | 23 dBm |
| Packet Size | 190 B | Carrier Freq. | 5860 MHz |
| $1/T_{Gen}$ | 10 Hz | Bandwidth | 10 MHz |
| MCS Index | 5 | Tx Antenna | 1 |
| $\{T_1, T_2\}$ | {1, 100} | Rx Antenna | 2 |
| PSSCH RBs | 20 | Antenna Height | 1.5 m |
| $P_{rsvp}$ | 100 ms | Antenna Gain | 3 dBi |
| $P_{step}$ | 100 ms | Noise Figure | 9 dB |

The receiver module in the simulator acquires the SINR for each received packet from the PHY layer and maps this value to the transport block error rate (TBLER) for the corresponding received packet. The SINR⇔TBLER mapping is known as the receiver model and is preset in the receiver module of the simulator. In order to decide whether a received packet can be demodulated correctly or not, the receiver module compares the corresponding BLER value with a random variable which is uniformly distributed between 0 and 1; the packet is considered as a failed reception if the BLER is greater than the random variable.

There exist a few receiver models for the sidelink mode-4 communication suggested in the literature such as [22] and [25]; We employed the model derived by Huawei in [26] which has the best fit to our simulation assumptions. It should be noted that many parameters, such as the number of re-transmissions, MCS, message size, and vehicle speed, are involved in the receiver error model. Thus, separate models should be used for different sets of parameters.

### C. Application Layer

We conducted our studies on a moving platoon of vehicles in a straight highway where vehicles were uniformly arranged in 4 lanes in the stretch of 2 km and with 3 m inter-lane spacing. We extracted the results from the middle 1000 m stretch in order to avoid the edge effect due to the discontinuity in the vehicle platoon edges. All vehicles were set to transmit BSMs with fixed power and message rate and form a fully connected network. Transceivers were assumed to operate in the half-duplex mode which means that a vehicle is not able to receive any data while it is in the transmit mode.

The message set dictionary in [6] states that vehicles broadcast their position, speed, and heading in BSMs every $T_{Gen}$ milliseconds in two digested and full-certificate versions which contain 190 and 300 bytes of data, respectively In our study we only consider the digested packet size to obtain a more meaningful analysis. We set the total transmit power to $P_{Tx} = 23$ dBm as it is allowed by [15] and utilize adjacent PSSCH and PSCCH sub-channelization scheme. Other configured simulation parameters are listed in Table III.

## IV. ANALYSIS AND RESULTS

In this section, we present an investigation of the earlier discussed SB-SPS mechanism and the impact of the parameter tuning on its performance. We start with a concise observation on the MCS index as it is defined in [3], where MCS={0, 1, …, 20} indices map to the modulation order ($Q$), i.e, number of bits per symbol, as well as the transport block size. We use this mapping to calculate the effective code rate for a safety beacon transmission using different MCS indices as it is illustrated in Figure 3.

The effective code rate (CR$_{eff}$) can be expressed as follows

$$\text{CR}_{\text{eff}} = \frac{TB_{size} \text{ (bit)}}{Q\left(\frac{bit}{Symbol}\right) \times 9\left(\frac{symbol}{subcarrier}\right) \times 12\left(\frac{subcarrier}{PRB}\right) \times N_{PSSCH}^{RB}} \quad (3)$$

where $TB_{size}$ is the transport block size defined in [3]. It should be noted that the current 3GPP standard does not support the modulation orders higher than $Q = 4$ (16-QAM modulation). One can observe that the maximum achievable code rate for the QPSK modulation is $\text{CR}_{\text{limit}} \cong 0.8$.

Since the network "reliability" and communication "latency" have paramount importance in the vehicular safety applications domain, we focus on two related metrics for the performance evaluation. However, the same metrics can be used as key performance indicators in general case. We utilize the packet error rate (PER) as an indicator of the network reliability and inter-packet gap (IPG) to evaluate the communication latency. It worth mentioning that a packet-drop between two close UEs may cause a safety threat and has much higher importance than a lost packet between two far UEs. Hence, illustrating PER as a function of transmitter-receiver distance will be more enlightening than only measuring its value for the whole network.

We consider four test scenarios with different congestion levels and vehicle speeds based on the suggested evaluation scenarios in [14]. Scenarios #1 to #4 relate to 12.5, 25, 50, and 100 (vehicles.km$^{-1}$.lane$^{-1}$) vehicle densities and 140, 70, 60, and 15 (km.h$^{-1}$) speeds, respectively. Scenarios #1 and #2 can be considered as freeway mobility models, in contrast, more congested and lower speed scenarios #3 and #4 simulate an urban mobility model. Considering the packet size and MCS index chosen in section III and Figure 3, it can be inferred that every 100ms transmission time interval (TTI) contains 200 radio resources (100 subframes and 2 sub-channels per subframe) which means that with an ideal centralized resource allocation and full-duplex operation, maximum 200 UEs (scenario #2) can fit into one TTI without experiencing packet collisions. However, in our case, nodes operate in half-duplex mode and utilize distributed resource allocation. Figure 4 shows the PER values over a range of transmitter-receiver distances. It can be observed that in the low-density case (scenario #1), the network maintains the 90% reliability level for almost the whole communication range while in the more congested cases, performance degeneration is noticeable.

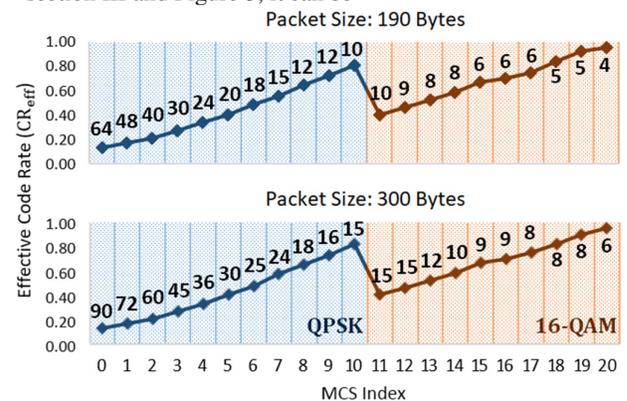

Fig. 3. Effective code rate values using different MCS indices for 190B (top) and 300B packet (bottom). Numbers on the curve denote $N_{PSSCH}^{RB}$.

For the vehicle density values higher than the network saturation point, where the number of available radio resources is less than the number of UEs, the network is no longer able to maintain the required packet delivery rate and PER increases significantly. This issue is due to the insufficient radio resources for all transmitting UEs and can be addressed by either limiting the transmit range of UEs, through decreasing the transmit power, or slowing down the beacon generation rate. However, such congestion control algorithm has not been specifically mandated by the 3GPP standard. It should be noted that the picking points in PER curves in Figure 4, are side-effects of the geographical conditions of the Fowlerville test track and become more significant in more congested scenarios due to the lower link budget. This anomaly is independent of the vehicular scenario and can be considered as a characteristic of the air interface.

It is also important to study the network behavior under the high loads in terms of communication latency. From an application layer point of view, one may be interested in measuring the time separation between two consecutive packet receptions, i.e., IPG. In the absence of transmission errors, every receiver should be able to receive a safety beacon from its neighboring UEs every 100ms. Packet drop nonetheless may result in much higher IPG values. Figure 5 shows the normalized histogram of IPG for the earlier discussed congestion scenarios. It can be noticed that in the densities below the saturation level, the network almost maintains the ~100ms IPG, however, increase in the vehicle density not only spreads the distribution of the IPG but also can cause an over 1s gap between two consecutive packet receptions. The wider distribution in scenario 3 and 4 can be interpreted as more unpredictability and hence less reliability in the network.

We also investigate the impact of two tunable parameters of the SB-SPS mechanism on its performance in terms of IPG and PER metrics. As it is discussed in section II, SB-SPS decreases the packet collision probability through removing the likely-to-collide radio resources from the candidate set. There are two contributing factors in this process; CSRs with RSRP higher than $Th_{SPS}$ are exempted from the candidate set and the 20% of the remainder list with the lowest $E_{x,y}$ are extracted. Hence, depending on the network congestion level, both $Th_{SPS}$ or the 20% requirement can become the limiting factor. In low vehicle densities, the amount of the exempted radio resources due to $Th_{SPS}$ is much less than 80% of the candidate set size, hence, this threshold does not have a significant effect in the SB-SPS process in such scenarios and the 20% requirement is playing the main role.

On the other hand, in highly congested scenarios, a big portion of the candidate set resources get exempted due to the $Th_{SPS}$ in a way that the remainder set does not satisfy the 20% requirement and SB-SPS increases this threshold value until the 20% condition is met and the threshold value is not the limiting factor in this case as well. However, it should be noted that our observations are in a fully-connected and low-dynamics network while that will not necessarily be the case in a real-world V2X communication network where hidden nodes, near-far effect, and highly dynamic conditions have a significant effect on the network performance. The $Th_{SPS}$ value, in fact, defines the maximum range that a vehicle should sense which means that information from UEs above that distance has no significant importance for the receiver.

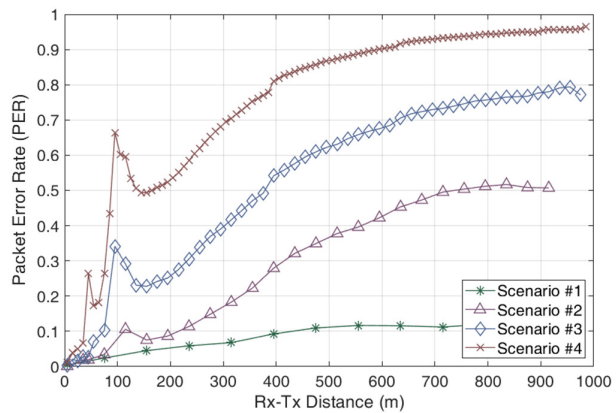

Fig. 4. Packet error rate (PER) curves in 4 congestion scenarios ($P_{resel}$=0.0, $Th_{SPS}$=−80dBm).

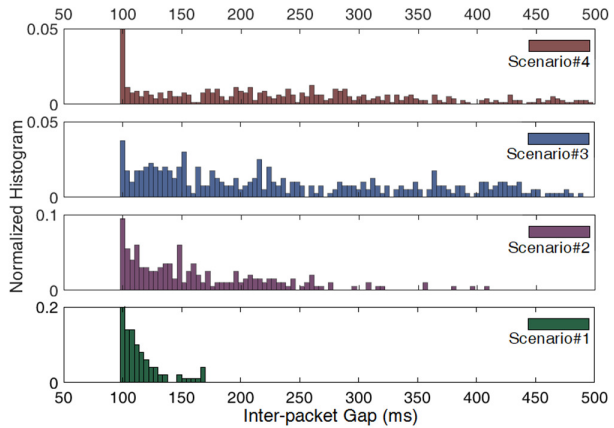

Fig. 5. Normalized histogram of inter-packet gap (IPG) in 4 congestion scenarios ($P_{resel}$=0.0, $Th_{SPS}$=−80dBm). Data points with inter-packet gap higher than 500 ms are not shown.

As it is stated in section II, when the resource re-selection counter (SLRRC) of a UE reaches to zero, it selects new resources with the probability $P_{resel}$. Figures 6 demonstrate the significance of this parameter on both IPG and PER metrics and it can be concluded that the higher probability of keeping previous resource selection by UEs leads to a more predictable and stable network, as if a UE keeps its previously selected resources, its neighbors have already "sensed" that selection during their sensing window period and will avoid selecting the same radio resources. In contrast, if the UE selects new resources after every counter, neighboring UEs will not have the chance to avoid collisions using their memory of the network. However, it should be noted that the behavior studied above is under the medium network congestion level, i.e., scenario #2, where deadlock loops such as two UEs keep colliding for a long period of time without being able to sense each other's transmissions, are less probable. Thus, tuning the $P_\text{resel}$ parameter will not be as straight-forward as discussed above in highly congested networks. Figure 7 shows the effect of $P_\text{resel}$ on the IPG in more detail. The lower probability of keeping the previously selected resources, i.e., higher $P_\text{resel}$, leads to widely spread distribution for IPG and significantly higher mean and median values. Data points with inter-packet gap higher than 350 ms are eliminated in Figure 7 for the sake of a more intelligible illustration.

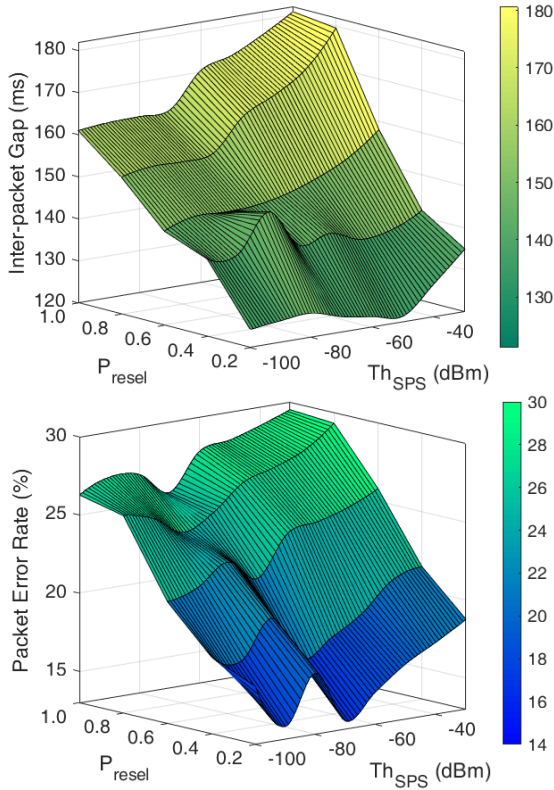

Fig. 6. Impact of $P_\text{resel}$ and $Th_\text{SPS}$ parameters on IPG and PER (Density=25 vehicle.$\text{km}^{-1}$.$\text{lane}^{-1}$).

The 3GPP standard mandates a distributed synchronization source such as GNSS (Global Navigation Satellite System) which can be utilized by UEs to provide the time-synchronization among UEs and enable them to transmit on synchronized subframes. However, there is no obligation by the current standard on the synchronized packet generation which means that every UE decides when to generate its safety beacons. It worth studying the effect of packet-generation synchronization among UEs as it has been proposed in some 3GPP workgroup discussions related to the geo-zoning [1] functionality [28]. We demonstrate results from three cases in Figure 8 and investigate the network behavior in those cases in order to understand the underlying phenomena.

In the synchronized case, all UEs generate their safety beacons in a synchronized fashion and repeat the transmission every 100 ms, in the second case we add a uniformly distributed random time offset between 0 to 99 ms to the first packet generation time of UEs and third case limits the time offset to 49 ms. Figure 8 shows that in the low-density scenario #1, time-offset in packet-generation time between the UEs have a noticeable impact, while this impact vanishes in the higher density scenario #2. Moreover, when the random time-offset value is limited to 49 ms, i.e., 49 subframes, the network shows lower PER values comparing to the synchronized and time-offset< 99 ms cases.

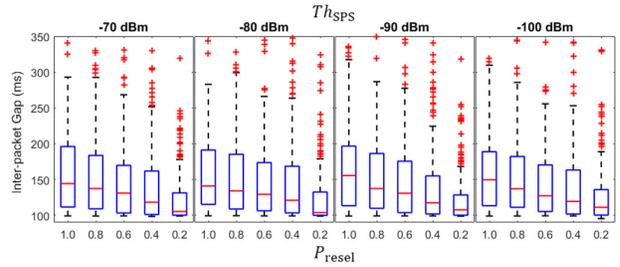

Fig. 7. Impact of $P_\text{resel}$ and $Th_\text{SPS}$ on the distribution of the inter-packet gap (IPG) values. (Density=25 vehicle.$\text{km}^{-1}$.$\text{lane}^{-1}$). Plot tails show the 95 percentile of the data.

Figure 9 helps us to understand the effect of the packet generation time offset among UEs. For the sake of simplicity, consider 2 vehicles, e.g., UEa and UEb, initiate their re-selection procedure at time $t_a$ and $t_b$ respectively, where $|t_b - t_a| = T_\text{offset}$. UEa selects its resource in the $(t_a + T_1, t_a + T_2)$ timespan while UEb does the same in the $(t_b + T_1, t_b + T_2)$ timespan, where $T_1$ and $T_2$ have been defined in section II. For a packet collision to happen it is required to have both UEs to select their radio resources from the intersection of their candidate sets, or in other words, from the $(t_b + T_1, t_a + T_2)$ period, as it is shown in Figure 9. It can be clearly affirmed that more overlap between the candidate sets of the UEs results in higher probability of collision. In the synchronized packet-generation case, the maximum overlap leads to a higher PER value and adding a time-offset to UEs packet-generation will decrease the intersection area.

However, offset values greater than $\lfloor (T_2 - T_1)/2 \rfloor$ will cause a greater overlap during the next period. It can be stated that the effective offset value can be expressed as $T_\text{offset}^\text{eff} = \max(T_\text{offset}, T_2 - T_\text{offset})$. In our case, we have set $T_1 = 1$ and $T_2 = 100$ which describes the behavior that has been observed in Figure 8. It can be concluded that mandating UEs to generate their safety beacons at the same time, using the GNSS clock, may lead to more packet collisions and consequently performance degradation.

## V. CONCLUDING REMARKS

C-V2X communication is being considered as a strong alternative for the vehicular communication technologies such as DSRC. This paper studies the reliability and latency of C-V2X in congested vehicular scenarios. The presented simulation results and analyses demonstrate the significance of the parameter tuning in the resource allocation mechanism to maximize performance and reliability in high-density networks. On the other hand, from the vehicular application perspective, a unified configuration has to be adopted by all vehicles in order to ensure communication reliability. We conclude that the introduced parameters have to be regulated for all vehicles by application layer standards, similar to what has been defined for DSRC in SAE J2945/1 standard. New workgroups are formed by the Society of the Automotive Engineers (SAE) to develop such standards for C-V2X; the current work progress is filed as SAE J3161 standard.

---

[1] Geo-zoning is a mechanism proposed by 3GPP which limits a UE's resource pool to specific resources based on its geographical location; it can be utilized in order to mitigate the packet collisions in highly congested scenarios and especially avoid the near-far problem.

For future research, we plan to investigate possible congestion control approaches in order to address scalability issues and satisfy the safety-related requirements. Furthermore, examining the impact of other configurable parameters, such as MCS index, report window size, and reservation interval, can assist rule-makers in specifying the mandatory system requirements of C-V2X.

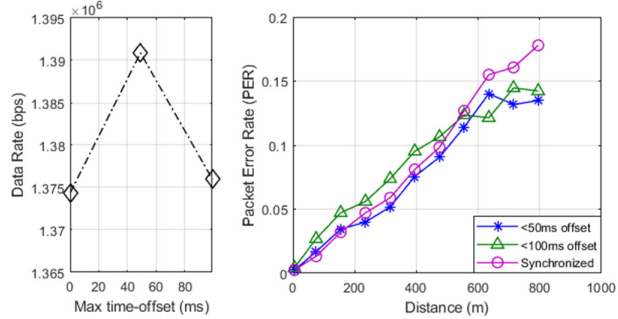

Fig. 8. Effect of packet generation synchronization on the network reliability in terms of the packet error rate (PER) and total data rate.

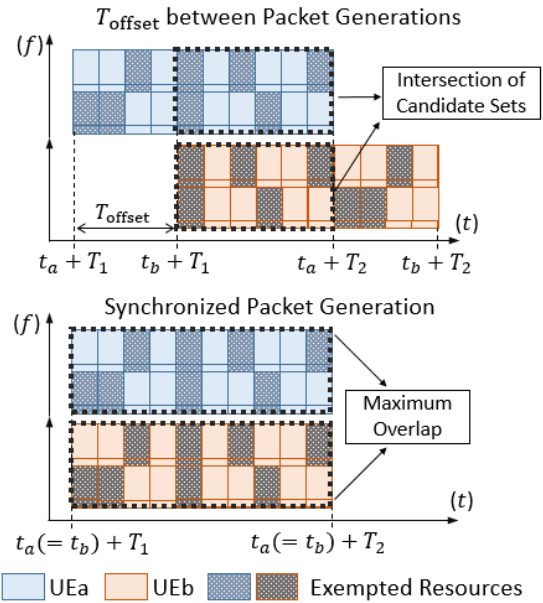

Fig. 9. Impact of the packet generation time-offset on collision probability for two given UEs.